Image definition evaluation functions for X-ray crystallography: A new perspective on the phase problem

Hui LI*, Meng HE* and Ze ZHANG

Beijing University of Technology, Beijing 100124, People's Republic of China

CAS Key Laboratory of Nanosystem and Hierarchical Fabrication, National Center for Nanoscience and Technology, Beijing 100190, People's Republic of China

Zhejiang University, Hangzhou 310014, People's Republic of China

Correspondence e-mail: huilicn@yahoo.com, mhe@nanoctr.cn

In memoriam of Ms. Guang-Lian Tian

**Abstract**

The core theme of X-ray crystallography is reconstructing the electron density distribution of crystals under the constraints of observed diffraction data. Nevertheless, the reconstruction of electron density distribution by straightforward Fourier synthesis is usually hindered due to the well-known phase problem and finite resolution of diffraction data. In analogy with optical imaging system, the reconstructed electron density map may be regarded as the image of the real electron density distribution in crystals. Inspired by image definition evaluation functions applied in auto-focusing process, we proposed two evaluation functions for the reconstructed electron density images. One of them is based on atomicity of electron density distribution and properties of Fourier synthesis. Tests were performed on synthetic data of known structures, and it was found that this evaluation function can distinguish the correctly reconstructed electron density image from wrong ones when diffraction data of atomic resolution is available. An algorithm was established based on this evaluation function and applied in reconstructing the electron density image from the synthetic data of known structures. The other evaluation function, which is based on the positivity of electron density and constrained power spectrum entropy maximization, was designed for cases where only diffraction data of rather limited resolution is available. Tests on the synthetic data indicate that this evaluation function may identify the correct phase set even for a dataset at the resolution as low as 3.5 Å. Though no algorithm of structure solution has been figured out based on the latter function, the results presented here provide a new perspective on the phase problem.

**Introduction**

It is interesting to compare X-ray crystallography with the optical imaging process (Miao *et al.*, 2015). In an optical imaging system, an object is illuminated by visible light, and the scattered light is then combined by the lenses to form an image which represent the object. In the case of X-ray crystallography, a crystal is illuminated by X-ray beam and the diffraction data is then measured directly since no appropriate "lens" is available for X-ray. One has to reconstruct the electron density image of the crystal from the diffraction data, which has been the core theme of X-ray crystallography since its advent. Theoretically, the electron density image of a crystal is just the inverse Fourier synthesis of its structure factors. Unfortunately, phases of structure factors can rarely recorded in diffraction experiments, and usually, only amplitudes of structure factor can be deduced from the diffraction data. Therefore, the electron density distribution of crystals cannot be reconstructed by a straightforward Fourier synthesis. Even worse, the phases "lost" in the diffraction experiments are more important than amplitudes of structure factors in defining the electron density distribution (Lattman and Derosier, 2008). This is so-called "phase problem" of X-ray crystallography.

Fourier synthesis can be made after assigning an arbitrary phase set to the observed moduli, but the resultant Fourier map generally will not represent the real electron density distribution in the crystal. If all possible phase sets for the observed moduli can be tested in a limited time, the correct phase set and subsequently the electron density image which represents the crystal must be included in these tests. Then the "phase problem" is converted to a question how to pick out the correct electron density image among all the possible Fourier synthesis maps. Such a question is reminiscent of the auto-focusing process which is extensively performed in optical imaging devices. In the auto-focusing process, the image definition evaluation functions are calculated in real time and focusing is then tuned accordingly. The image definition evaluation functions describe the properties of photographs mathematically (Chen *et al.*, 2013) and the properly focused image is then picked out based on the evaluation functions. Inspired by this, we try to establish evaluation functions for electron density image reconstructed from diffraction data. Hopefully, the evaluation function can pick out the electron density image which represents the crystal. For this purpose, the evaluation function should describe the general characteristic of electron density distribution in real crystals, and varies sensitively with phase sets assigned to the moduli to distinguish the faithfully reconstructed electron density distribution from the artifacts.

The commonly recognized characteristics of electron density in crystals are atomicity and positivity. Actually, the atomicity and positivity have been extensively utilized in many (if not all) methods for structure solutions by X-ray diffraction. For example, the Patterson method (Patterson, 1934) and the recently popular charge flipping algorithm (Oszlányi and A. Sütő, 2004; 2008) are based on atomicity. Atomicity and positivity are also the prerequisite of the direct method (Woolfson, 1987; Woolfson and Fan, 1995), though apparently it works in reciprocal space. The success of the Patterson method, the direct method and the charge flipping algorithm indicates that electron density images can be faithfully reconstructed by combining the constraints of atomicity and/or positivity and observed structure factor moduli, at least in some cases.

Information theory provides another constrains on the reconstructed electron density image: the entropy of the reconstructed electron density distribution should be maximized under the constraints of observed structure factor moduli. An iterative procedure has been established based on the constrained entropy maximization for reconstructing electron density distribution from X-ray diffraction data (Collins, 1982; Sakata and Sato, 1990). This algorithm is usually referred to as the maximum entropy method (MEM) (Wu, 1997). The bound of positivity is usually maintained by the electron density distribution reconstructed with the MEM.

In this study, we try to construct image definition evaluation functions for electron density image based on atomicity, positivity and constrained power spectrum entropy maximization, the validity of which in structure solutions has been confirmed by the practices in past decades. Two evaluation functions are proposed. One of them is based on atomicity, applicable for cases where diffraction data of atomic resolution is available; the other combines the bound of positivity and constrained power spectrum entropy maximization, aiming for cases where only diffraction data of rather limited resolution is available.

**I.    Image definition of electron density image and phases of structure factors**

The concept of image definition has been accepted extensively but its accurate definition is still under controversy. We prefer to define the image definition as a quantity which measures how well an image represents the object. The more information about the object is included in an image, the better the image definition. Many image definition evaluation functions have been proposed to assess the definition of an image quantitatively. These evaluation functions are calculated based the image itself.

In comparison with the image definition of a photograph, the image definition of a reconstructed electron density image seems to be more elusive, possibly because that the real electron density distribution of a crystal cannot be viewed directly by eyes. The pictures shown in Figure 1 may be helpful to understand the concept of image definition of a reconstructed electron density image. The electron density image of GaN reconstructed by combining the structure factor moduli (up to 0.8 Å) and correct phases represents the structure well (Figure 1a). With more and more phases of structure factors being replaced by random values, the information on the structure contained in the reconstructed electron density images decrease gradually. In the electron density image reconstructed with 25% random phases, the positions of Nitrogen atoms can hardly be identified (Figure 1b). When 50% or 75% phases are replaced by random values, only the positions of Gallium atoms can be discerned (Figure 1c and d). In the electron density image reconstructed with all random phases (Figure 1e), no any information on the structure is available. The information on the structure carried by the reconstructed electron density image decreases with the increasing amount of phases being replaced by random values, resulting in worse image definition. The example shown in Figure 1 not only illustrates the concept of image definition of electron density images, but also demonstrates that the image definition varies with phases of structure factors, and best image definition is achieved when correct phases are applied.

**II.  Image definition evaluation function for diffraction data of atomic resolution**

1. Construction of image definition evaluation function

As mentioned above, many image definition evaluation functions have been established to assess the quality of photographs. An important kind of evaluation functions are based on the idea that the more high frequency information contained in the image, the better the image definition is. The energy of high frequency information of an image is used to measure the image definition. However, this idea cannot be applied directly in the evaluation of the reconstructed electron density image of crystals, because all the Fourier synthesis maps created by combing fixed structure factor moduli with various phase sets will have the same energy of high frequency information, which depends solely on the moduli of structure factors. Hence, evaluation functions for the reconstructed electron density images have to be based on other intrinsic characteristic. As illustrated in Figure 1, in a correctly reconstructed electron density image, most electrons are concentrated at the positions of atoms while only few electrons distribute in the intermediate region between atoms. This is well-known atomicity of electron density distribution. On the contrary, when random phases are assigned to the structure factor moduli, the resultant electron density image shows a much evener distribution of electrons. It implies that the extent to which the electrons are concentrated can be used to evaluate the image definition of the reconstructed electron density image. The ratio of the sum of electron density at the atomic region to that in the intermediate region seems to be a good indicator to measure the extent to which the electrons are concentrated. Unfortunately, this ratio cannot be calculated before the structure is determined because one cannot discern the atomic and intermediate regions. An alternative way is to set a slightly positive threshold of electron density $\rho_t$, and use a quantity such as $\Sigma \rho_i^*$ (where $\rho_i^* > \rho_t$) to measure the extent to which the electrons are concentrated in the reconstructed electron density images. Nevertheless, the threshold should be structure dependent, and it might be difficult to give a reasonable estimate of $\rho_t$.

Due to the truncation effect of Fourier synthesis, both positive and negative electron density will be observed in the reconstructed electron density image. In a correctly reconstructed image, high positive electron density will appear at the position of atoms while in the intermediate region between atoms there are both small positive and negative electron densities. According to the property of Fourier transform, for any electron density image, the sum of positive density is equal to that of negative density when $F_{000}$ is not included in the Fourier synthesis. This means that all negative density lies in the intermediate region while most positive density is concentrated at the atomic positions. In terms of Parseval's theorem, for any given group of structure factor moduli we have

$$\sum_i \rho_+^2 + \sum_j \rho_-^2 = \text{constant}$$

where $\rho_+$ and $\rho_-$ represent non-negative and negative density, respectively, at a certain grid of electron density image. Parseval's theorem and a brief derivation of the above corollary are presented as the supporting information. As discussed above, for a correctly reconstructed electron density image, there is very little positive electron density in the intermediate region between atoms, so $\Sigma\rho_+^2$ is a good approximation of the sum of squared electron density at the atomic positions ($\Sigma\rho_{atom}^2$), while $\Sigma\rho_-^2$ dominates the sum of squared electron density in the intermediate regions between atoms ($\Sigma\rho_{inter}^2$). Hence, the ratio $\Sigma\rho_+^2/\Sigma\rho_-^2$ is a good approximation for $\Sigma\rho_{atom}^2/\Sigma\rho_{inter}^2$, which measures well the extent to which the electrons are concentrated. Thus, we establish the first image definition evaluation function for the reconstructed electron density images:

$$Tian1 \equiv \frac{\sum_i \rho_+^2}{\sum_j \rho_-^2}$$

2. Verification and limitation

Tests are performed on the synthetic diffraction data of tens known structures to verify the validation of Tian1 and find its limitations. The moduli of structure factors calculated from the known structure data are taken as the synthetic diffraction data. A series of Fourier synthesis maps are then created by combing these structure factor moduli with correct phases, all-zero phases and random phase sets, respectively. Evaluation function Tian1 is calculated for each Fourier synthesis map. Typical results of such tests are presented in Figure 2. As revealed by Figure 2, random phases will lead to a quite low Tian1 value, which is close to 1. This is reasonable because the random phases will lead to a random distribution of electron density, implying that statistically $\Sigma\rho_+^2$ is equal to $\Sigma\rho_-^2$. Both correct phases and all zero phases result in high Tian1 values, indicating that electrons are well concentrated in the reconstructed electron density images. When the data of atomic resolution (1.0 Å) is available, the correct phases give a higher Tian1 value than all zero phases do. When the resolution of the data is getting worse, the decreased Tian1 values are observed for both the correct and all zero phases. Unfortunately, the Tian1 value for the correct phase set decreases faster than that for all zero phases. At the resolution of 1.5 Å, the comparable Tian1 values are obtained for the correct phase set and all zero phases. This implies the invalidation of Tian1 in identifying the correctly reconstructed electron density image. The invalidation of Tian1 in cases where data of atomic resolution is not available can be well understood. In such cases, positive electron density will no longer be mainly concentrated at the positions of atoms in a correctly reconstructed electron density image. Significant positive electron density will appear in the intermediate regions between atoms. On the contrary, all zero phases always lead to the concentration of electron density at the origin of unit cell. The evaluation function Tian1 measures the extent to which the electron density is concentrated, so it works well only when the data of atomic resolution is available. The resolution limit for Tian1 might be structure dependent, but 1.5 Å~1.8 Å seems to be a reasonable estimate.

3. Algorithm for structure solution based on Tian1

An algorithm for structure solution has been established based on the evaluation function Tian1. The flow chart of the algorithm is given as follows:

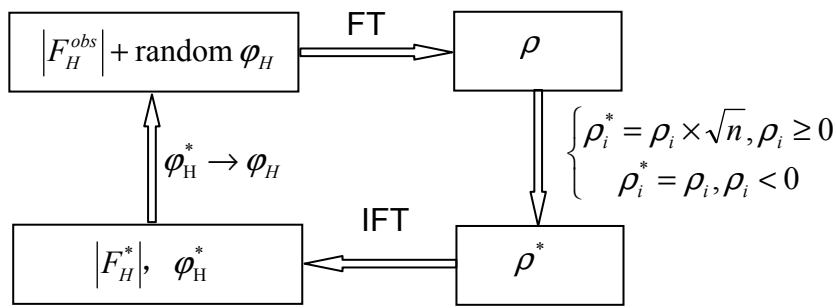

At the very beginning of the iterative process, a random phase is assigned to each observed structure factor modulus, and then an electron density image is calculated by Fourier synthesis. The original electron density image $\rho$ is subsequently modified by multiplying each non-negative electron density by $\sqrt{n}$ ($n > 1$) while keeping all negative electron densities unchanged. The purpose of electron density modification is to improve the Tian1 value of the original electron density image by $n$ times. The modified electron density image $\rho^*$ is then inversely Fourier transformed to generate new moduli and phases of structure factors. The latest phases are then combined with the observed structure factor moduli to calculate new electron density image. The iterative process is repeated until a maximized Tian1 value has been achieved.

The algorithm is tested on synthetic diffraction data of known structures at a resolution of 0.8 Å. In the tests, symmetry was not taken into account and all structures were solved in $P$1 space group. The resultant final electron density images obtained by this algorithm were compared with the known crystal structure by superimposing the electron density images upon atomic structure model using the software VESTA (Momma and Izumi, 2008). Electron density images are thought to be correct when they are consistent with atomic structure model. It is worth noting that atomic coordinates in the structure model usually have to be translational shifted and/or inverted in some cases to make a direct comparison with the electron density images. The largest structure has been tested with this algorithm is $K_{14}((H_2O)W_{19}P_2O_{69})(H_2O)_{24}$ (ICSD-66144), which contains 2064 non-hydrogen atoms in the unit cell including 112 K and 32 O undetermined. This structure was solved successfully in $P$1 space group with the algorithm presented here, and all positions of the non-hydrogen atoms in the known structure model can be identified in the final reconstructed electron density image.

This algorithm is somewhat similar to the auto-focusing process. In the auto-focusing process, the focusing is tuned to achieve the extremum of an image definition evaluation function. In the algorithm based on Tian1, the electron density image is modified to maximize the Tian1 value. The algorithm is a dual-space method, and the similarity between this algorithm and the charge flipping as well as error-reduction algorithms (Fienup, 1982; Stark, 1987) is apparent: In all cases, the electron density is modified iteratively to converge to an image which represents the crystal. Computer programs are also developed based on the charge flipping and error-reduction algorithms. All three algorithms are tested on the synthetic data of $Al(IO_3)_3(H_2O)_8$ (ICSD-200619) at a resolution of 0.8 Å. $Al(IO_3)_3(H_2O)_8$ crystallizes in $P1$ space group, and the lattice parameters are a = 7.473 Å, b = 8.392 Å, c = 13.574 Å, $\alpha$ = 89.93 °, $\beta$ = 87.62 °, $\gamma$ = 63.58 °. There are 42 non-hydrogen atoms in the unit cell of $Al(IO_3)_3(H_2O)_8$ and all hydrogen atoms are undetermined in the structure model. The structure can be solved with all three algorithms. The Tian1 value is calculated in each iteration cycle for all three algorithms and presented in Figure 3a. For the algorithm proposed here, the Tian1 value increases monotonously in the iteration process as expected. It is interesting to note that this value increases monotonously also in the error-reduction iteration process. It is slightly different in the case of the charge flipping algorithm. The value of Tian1 function increases rapidly in the iteration process and soon an extremum is reached. After that this value decreases a little, and then is maintained at a level close to the extremum. The extrema achieved by the error-reduction and our algorithm is slightly larger than the value corresponding to the Fourier synthesis maps created using the calculated structure factors. This indicates that the over-concentration of the electron density takes place in the iteration process. On the contrary, the extremum of Tian1 function reached by the charge flipping algorithm is significantly lower than the value corresponding to the theoretical Fourier synthesis map. This implies that electrons have not been sufficiently concentrated in the electron density image obtained with the charge flipping algorithm. Nevertheless, electron density images created with all three algorithms are good enough to identify the positions of non-hydrogen atoms, as demonstrated by Figure S1 in the supporting information.

Although the Tian1 value evolves similarly in the iterative process of all three algorithms, the algorithm based on the Tian1 value differs from the charge flipping and error reduction algorisms in that the goal of the iterative process is improving the Tian1 value for the former, while for the latters, the Tian1 value just increases unintentionally with the evolution of electron density distribution. The variation of the Tian1 value observed in the iterative process of the charge flipping and error reduction algorithms further corroborates the idea that the image definition of the reconstructed electron density images can be used to identify the correct phase assignment, and the Tian1 value is effective in evaluating the image definition of the electron density images reconstructed from atomic resolution data.

A preliminary check has been performed to evaluate the effect of the uncertainties of the data on the applicability of the algorithm based on Tian1. We modified each calculated structure factor modulus of $Al(IO_3)_3(H_2O)_8$ by multiplying it with $(1 + x \times 30\%)$ (where $x$ is a random value in the range from -1 to 1) to simulate the experimental diffraction data with random errors. The structure can be retrieved from such a dataset by the algorithm based on Tian1, the charge-flipping and error reduction algorithms. The evolution of Tian1 in the iterative process is similar to that observed in the case of error-free data, as shown in Figure 3b. With the introduction of random errors in the simulated data, the maximum Tian1 values obtained in the iterative process decrease slightly in comparison with those resulted from error-free data for all three algorithms. Interestingly, the structure is retrieved with less iteration cycles from the data with random errors.

**III. Image definition evaluation function for diffraction data of limited resolution**

1. Construction of image definition evaluation function

As discussed above, the evaluation function Tian1 is valid only for the diffraction dataset of atomic resolution. However, it is more desired to establish an evaluation function which is applicable for the electron density image reconstructed from the diffraction data at rather limited resolution. Methods of structure solution from the diffraction data at atomic resolution, such as the direct method, Patterson method and charge flipping algorithm, have been well developed. Nevertheless, in many cases, such as in protein crystallography, the diffraction data of atomic resolution is only rarely available. Usually, techniques of isomorphous replacement or anomalous dispersion have to be used to solve the phases. Anomalous dispersion requires the presence of heavy atoms in the structure, and the preparation of isomorphous heavy-atom derivative crystals is not trivial. Hence, the technique of reconstructing the electron density image directly from a diffraction dataset of low resolution is still strongly desired.

The positivity constraint on the reconstructed electron density image is independent on the resolution. Under the constraint of positivity, a correctly reconstructed electron density image should give the structure factors consisting with the observed data. For any phase set which is assigned to the observed moduli, the consistence between the calculated and "observed" structure factors can be measured by the residual factor, which is defined as

$$R_{mem} = \frac{1}{M}\sum \frac{(F_{obs}^{H} - F_{cal}^{H})^2}{(\sigma F_{obs}^{H})^2}$$

where $F_{obs}^{H}$ is the "observed" structure factor, $F_{cal}^{H}$ is the calculated structure factor, $\sigma F_{obs}^{H}$ is the standard deviation of $F_{obs}^{H}$, and M the number of structure factors. The "observed" structure factor is the combination of observed modulus and assigned phase. $F_{cal}^{H}$ is generated by calculating the Fourier transform of the electron density distribution.

In MEM calculations, the entropy of an electron density image is defined as

$$H = -\sum_{N}\left(\frac{\rho_i}{Z}\right)\ln\left(\frac{\rho_i}{Z}\right)$$

where $\rho_i$ is electron density at the $i$th grid of the unit cell, $Z$ is the sum of $\rho_i$ across the unit cell, and N is the number of grids in the unit cell. Reconstructing an electron density image with MEM is modifying the initial electron density distribution iteratively to reach a least $R_{mem}$ subject to maximization of the entropy.

It was found that the residual factor $R_{mem}$, which combines the constraints of positivity and observed data on the electron density image, is sufficient to identify the correct image when the high resolution diffraction data is available. When the resolution of diffraction data deteriorates, less constrains from the observed data are available on the reconstructed electron density image. Then iteration of MEM is more readily converged, yielding quite high entropy H and rather low $R_{mem}$. Unfortunately, in such cases, a low $R_{mem}$ is no longer a guarantee of the correctness of the reconstructed electron density image. Additional constraint has to be found to identify the correctly reconstructed image. The entropy H seems to be a good choice. Nevertheless, the entropies H of different electron density images reconstructed with MEM cannot be compared with each other directly since they are obtained under the constraint of various $R_{mem}$. Here we introduce the power spectrum entropy S, which is defined as

$$S = -\sum_N \left(\frac{\rho_i^2}{C}\right) \ln\left(\frac{\rho_i^2}{C}\right)$$

where C is the sum of $\rho_i^2$ across the unit cell. Different from the entropy H which is calculated from a positive definite electron density image, the power spectrum entropy S is based on a Fourier synthesis map. According to Parseval's theorem, for any given group of structure factor moduli, C is a constant. The power spectrum entropy S of a Fourier synthesis map may be taken as the entropy H of an image of $\rho_i^2$. Then the power spectrum entropy S measures the extent to which the image of $\rho_i^2$ is close to a flat distribution. Due to the correlation between the image of $\rho_i^2$ and the Fourier synthesis map, we suppose that S also measures the flatness of the Fourier synthesis map approximately. We construct an evaluation function by combing S and $R_{mem}$, which is defined as

$$Tian2 \equiv R_{mem} \times \exp(S_{ideal} - S)$$

where $S_{ideal} = \ln N$, N is the number of grids in the unit cell. $S_{ideal}$ is the power spectrum entropy of a completely flat electron density image. This evaluation function is based on the idea that a correct phase set will result in a positive definite electron density image consisting with the observed data and simultaneously a Fourier synthesis map which is as flat as possible.

2. Verification and limitation

Evaluation functions Tian2 is tested on the synthetic diffraction data of two known structures. One is $Al(IO_3)_3(H_2O)_8$, the crystallographic data of which have been given in the previous section. The other is $C_{252}H_{326}O_{19}$ which is reported by Czugler et al (2003). This compound crystallizes in a triclinic unit cell with lattice parameters $a$=16.909 Å, $b$=18.772 Å, $c$=21.346 Å, $\alpha$=111.46°, $\beta$=103.38°, $\gamma$=107.74°. Space group of both structures is $P$1. The unit cells of $Al(IO_3)_3(H_2O)_8$ and $C_{252}H_{326}O_{19}$ are divided into 64 × 64 × 128 and 170 × 190 × 220 electron density pixels, respectively. The moduli of structure factors calculated from the known structures are taken as synthetic diffraction data. For each synthetic dataset, a phase set is assigned, and then a positive definite electron density image is created with MEM. The value of $R_{mem}$ is recorded when the convergence of MEM iterations has been reached. A Fourier synthesis map is also calculated using the synthetic structure factor moduli and the assigned phase set. Subsequently, the power spectrum entropy $S$ of the Fourier synthesis map is calculated. The value of the evaluation function Tian2 is then calculated using $R_{mem}$ and $S$. For each synthetic dataset of $Al(IO_3)_3(H_2O)_8$, the value of Tian2 is calculated for the correct phase set, all-zero phase set and 15 random phase sets, respectively. For $C_{252}H_{326}O_{19}$, the value of Tian 2 is calculated for 10 random phase sets in addition to the correct and all-zero phase set. The results of tests are presented in Figure 4. As revealed by Figure 4, for high resolution data set, the value of Tian2 can distinguish the correct phase set from the all-zero and random phase sets readily. With the deterioration of data resolution, the difference among the values of Tian2 resulting from correct, all-zero and random phase sets decreases gradually. However, up to the resolution of 3.5 Å, the evaluation function Tian2 can still distinguish the correct phase set from the wrong ones for the structure of $C_{252}H_{326}O_{19}$. Based on the very limited tests, the evaluation function Tian2 is expected to be promising in cases where only diffraction dataset of rather limited resolution is available.

3. Algorithm for structure solution based on $R_{mem}$

Algorithms have not been figured out yet to solve structures based on the evaluation function Tian2. Nevertheless, an algorithm based on $R_{mem}$ is developed to solve structures from high resolution data. The scheme of the algorithm is shown below:

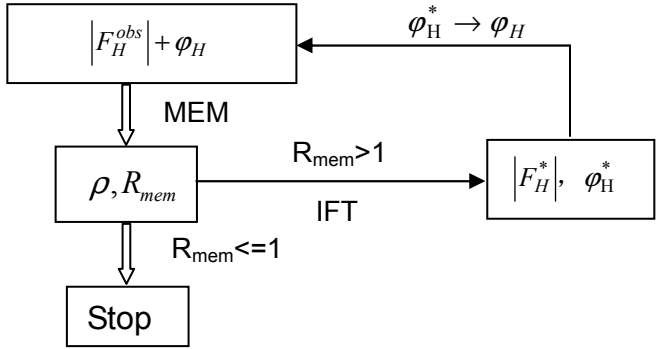

The positive definite electron density image ρ is first created by MEM using the observed moduli of structure factors $|F_H^{obs}|$ and the randomly assigned phases $\varphi_H$, $R_{mem}$ is calculated simultaneously. If the iterative MEM process results in an $R_{mem}$ which is less than 1, the electron density image $\rho$ is then considered to be correct. If the iterative MEM process results in an electron density image $\rho$ with $R_{mem} > 1$, then $\rho$ is inversely Fourier transformed to generate new moduli $|F_H^*|$ and phases $\varphi_H^*$ of structure factors. Replacing $\varphi_H$ with the latest $\varphi_H^*$ and then start the new iterative MEM process. The whole process is terminated when an $R_{mem} < 1$ is achieved.

The algorithm is tested on several synthetic datasets at the resolution of 0.8 Å. It is demonstrated that the correct electron density image can be reconstructed from high resolution data under the constraint of $R_{mem} < 1$.

## IV. Discussions and concluding remarks

In this contribution, the solution to the phase problem of X-ray crystallography is attempted from a new perspective of image definition evaluation functions. Two evaluation functions are established to assess the image definition of the electron density images reconstructed from the X-ray diffraction data. The first function, Tian1, can be used to identify the image constructed with the correct phase set when the atomic resolution data is available. The other function, Tian2, is still applicable when the resolution of dataset is as low as 3.5 Å. Therefore, it is expected that the function Tian2 may find applications in protein crystallography.

An iterative algorithm based on the function Tian1 has been established to solve structures from the atomic resolution data. Algorithms of structure solution based on the function Tian2 have not been figured out yet. Global optimization methods are being considered in designing the algorithms based on Tian2. Introducing direct method into the potential algorithms based on Tian2 may improve the efficiency of structure solution.

There are perhaps more appropriate functions to evaluate the image definition of the reconstructed electron density image, and more efficient algorithms for structure solution from X-ray diffraction data. For example, we are now developing another two modified Tian's definition functions. One is $Tian1a \equiv Tian1 \times \exp(S)$ for high resolution data. The other is $Tian2a = Tian2 \times \rho_{max}$, where $\rho_{max}$ is the maximum value of the charge density, for general case. Nevertheless, the significance of this work lies in that it provides a new perspective on the phase problem of X-ray crystallography.

The work was financially supported by the National Natural Science Foundation of China (Grant No. 11474281 and 51272049).

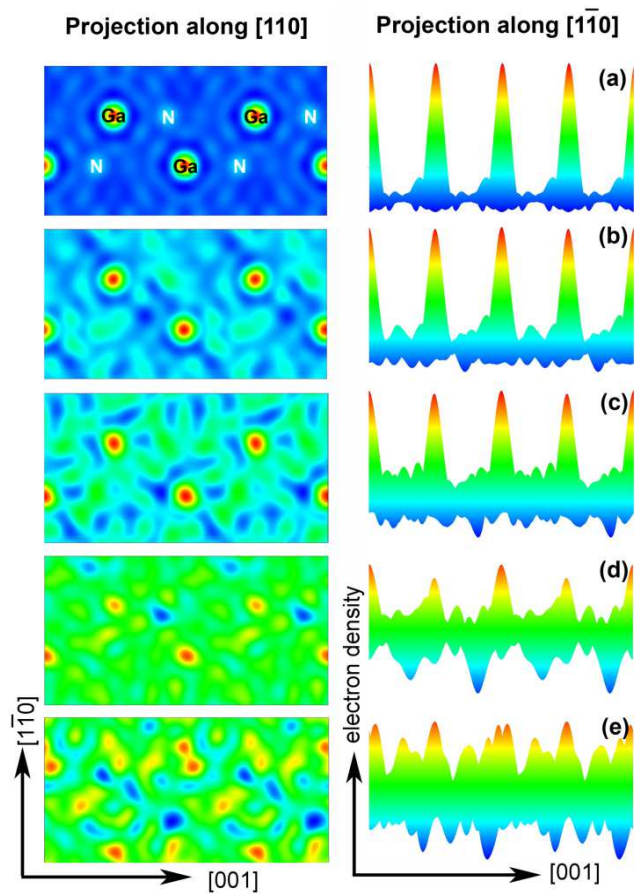

Figure 1. Electron density distribution on (110) plane of hexagonal GaN calculated by Fourier synthesis using synthetic structure factor moduli and (a) correct phases, (b) 25% random phases, (c) 50% random phases, (d) 75% random phases and (e) 100% random phases. The resolution of the synthetic diffraction data is 0.8 Å.

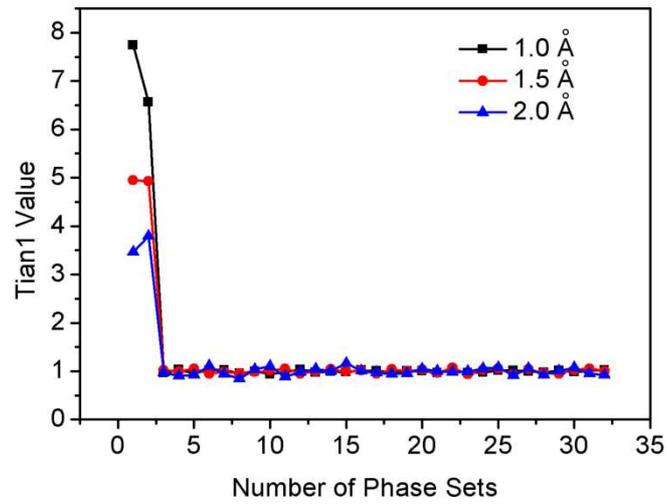

Figure 2. Evaluation function Tian1 of electron density images created by Fourier synthesis using synthetic structure factor moduli of $Al(IO_3)_3(H_2O)_8$ and correct phases, all zero phases or random phase, respectively. The first phase set consists of correct phases; the second one is made up of all zero phases, and all other phase sets are totally random. The unit cell is divided into $64 \times 64 \times 128$ grids.

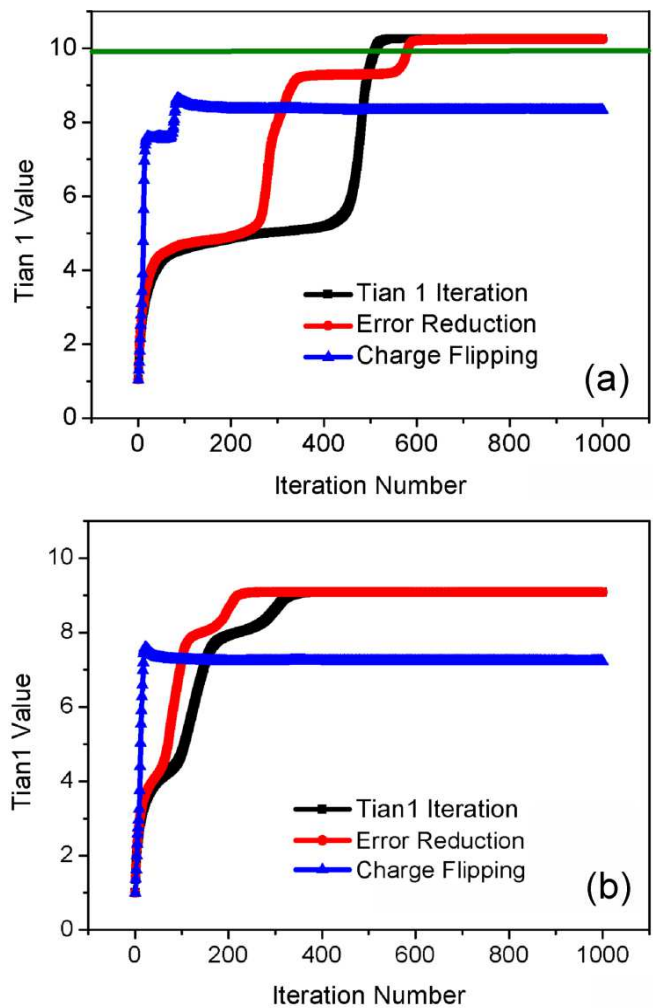

Figure 3. The evolution of evaluation function Tian1 in the iterative cycles of structure solution of $Al(IO_3)_3(H_2O)_8$ from the simulated error-free data (a) and the simulated data with random uncertainties (b). The green horizontal line in (a) indicates the Tian1 value of the Fourier synthesis map created using calculated structure factors. The resolution of synthetic diffraction data is 0.8 Å.

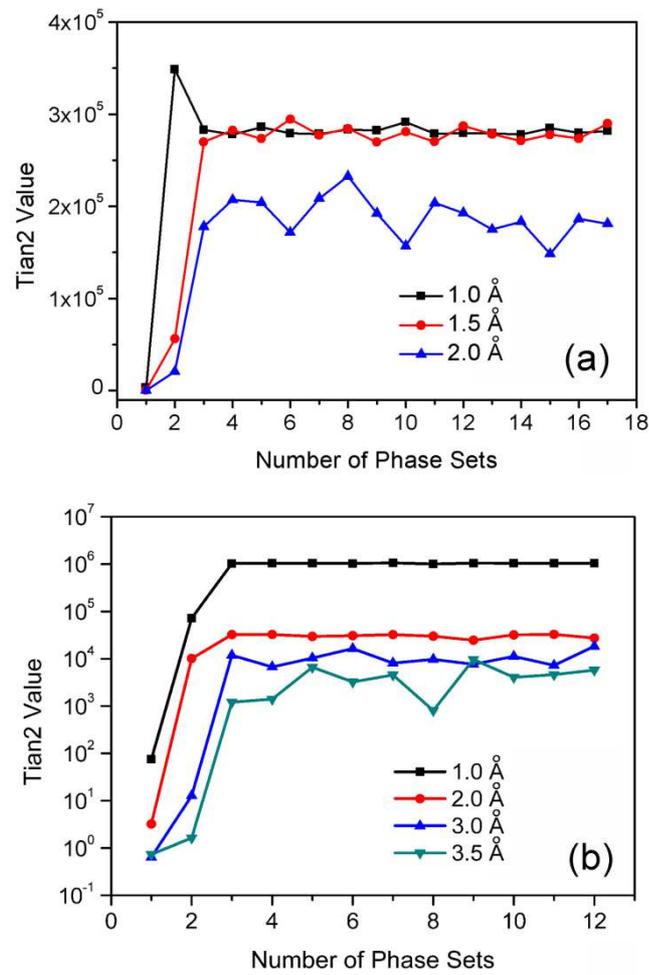

Figure 4. Evaluation function Tian2 of electron density images reconstructed using synthetic structure factor moduli and correct, all zero and random phase sets, respectively. The first phase set consists of correct phases; the second one is made up of all zero phases, and all other phase sets are totally random. (a) $Al(IO_3)_3(H_2O)_8$, (b) $C_{252}H_{326}O_{19}$. The unit cells of $Al(IO_3)_3(H_2O)_8$ and $C_{252}H_{326}O_{19}$ are divided into $64 \times 64 \times 128$ and $170 \times 190 \times 220$ grids, respectively.

# Supporting Information

Image definition evaluation functions for X-ray crystallography: A new perspective on the phase problem

Hui LI*, Meng HE* and Ze ZHANG

Beijing University of Technology, Beijing 100124, People's Republic of China

CAS Key Laboratory of Nanosystem and Hierarchical Fabrication, National Center for Nanoscience and Technology, Beijing 100190, People's Republic of China

Zhejiang University, Hangzhou 310014, People's Republic of China

Correspondence e-mail: huilicn@yahoo.com, mhe@nanoctr.cn

In memoriam of Ms. Guang-Lian Tian

## 1. Parseval's theorem and corollaries for electron density images reconstructed by Fourier synthesis.

Parseval's theorem:

Suppose that $A(x)$ and $B(x)$ are two square integrable complex-valued functions of period $2\pi$ with Fourier series

$$A(x) = \sum_{n=-\infty}^{\infty} a_n \exp(inx)$$

and

$$B(x) = \sum_{n=-\infty}^{\infty} b_n \exp(inx)$$

respectively. Then

$$\sum_{n=-\infty}^{\infty} a_n \overline{b_n} = \frac{1}{2\pi} \int_{-\pi}^{\pi} A(x) \overline{B(x)} dx$$

where $i$ is the imaginary unit and horizontal bars indicate complex conjugation.

The structure factor $F(H)$ is the Fourier transform of the electron density distribution $\rho(r)$,

$$F(H) = \sum_{r=0}^{n} \rho(r) \exp(-2\pi i H \cdot r)$$

then

$$\overline{F(H)} = \sum_{r=0}^{n} \overline{\rho(r)} \exp(2\pi i H \cdot r)$$

where $\overline{F(H)}$ and $\overline{\rho(r)}$ are complex conjugate functions of $F(H)$ and $\rho(r)$, respectively.

According to Parseval's theorem, we have

$$\frac{1}{2H} \int_{-H}^{H} F(H)\overline{F(H)} dH = \sum_{r} \rho(r)\overline{\rho(r)}$$

Because $\overline{\rho(r)} = \rho(r)$ ($\rho(r)$ is a real function), then

$$\int_{-H}^{H} |F(H)|^2 dH = 2H \times \sum_{r} \rho(r)^2$$

Since $F(H)$ is discrete, the above equation can be re-written as

$$\sum_{H} |F(H)|^2 = M \sum_{r} \rho(r)^2$$

where M is the number of structure factors.

For any given group of $|F(H)|^2$ (diffraction data), $\sum_{H} |F(H)|^2$ is a constant, then $\sum_{r} \rho(r)^2$ is also a constant.

Because

$$\sum \rho_+^2 + \sum \rho_-^2 = \sum_{r} \rho(r)^2$$

where $\rho_+$ and $\rho_-$ are the non-negative and negative electron density, respectively, at the grids across the unit cell, then

$$\sum \rho_+^2 + \sum \rho_-^2 = \text{constant}$$

## 2. Iterative process of reconstructing the electron density image with maximum entropy method (MEM).

The entropy of the electron density distribution is defined as

$$H = -\sum \left(\frac{\rho_i}{Z}\right) \ln\left(\frac{\rho_i}{Z}\right)$$

where $\rho_i$ is the number of electrons at the $i$th grid of the unit cell, Z is total number of electrons in the unit cell, $Z = \sum \rho_i$.

For any phase set which is assigned to the observed moduli, the consistence between the calculated and "observed" structure factors can be measured by the residual factor, which is defined as

$$R_{mem} = \frac{1}{M} \sum \frac{(F_{obs}^H - F_{cal}^H)^2}{(\sigma F_{obs}^H)^2}$$

where $F_{obs}^H$ is the "observed" structure factor, $F_{cal}^H$ is the calculated structure factor, $\sigma F_{obs}^H$ is the standard deviation of $F_{obs}^H$, and M is the number of structure factors. The "observed" structure factor is the combination of observed modulus and assigned phase. $F_{cal}^H$ is generated by calculating the Fourier transform of the electron density distribution.

Build a function

$$G(\rho) = H + \lambda R_{mem} + Z$$

where λ is a disposable constant to be evaluated. The goal of iterative process of MEM is to maximize the entropy $H$ subject to the conditions that $R_{mem}$ is minimized and Z remains unchanged. This was achieved by setting

$$\frac{\partial G}{\partial \rho_i} = \frac{\partial H}{\partial \rho_i} + \lambda \frac{\partial R_{mem}}{\partial \rho_i} + \frac{\partial Z}{\partial \rho_i} = 0$$

Because

$$\frac{\partial R_{mem}}{\partial \rho_i} = -2 \times \frac{1}{M} \sum \frac{(F_{obs}^H - F_{cal}^H)}{(\sigma F_{obs}^H)^2} \times \frac{\partial F_{cal}^H}{\partial \rho_i}$$

and

$$F_{cal}^H = \sum_{r=0}^{n} \rho(r) \exp(-2\pi i H \cdot r)$$

then

$$\frac{\partial R_{mem}}{\partial \rho_i} = -2 \times \frac{1}{M} \sum \frac{(F_{obs}^H - F_{cal}^H)}{(\sigma F_{obs}^H)^2} \exp(-2\pi i H \cdot r)$$

Since
$$\frac{\partial H}{\partial \rho_i} = \frac{(-\ln(\rho_i)+1)}{Z}$$

and
$$\frac{\partial Z}{\partial \rho_i} = 1$$

then
$$\frac{\partial G}{\partial \rho_i} = -\frac{\ln(\rho_i)}{Z} + \frac{1}{Z} - \frac{2\lambda}{M} \times \sum \frac{(F_{obs}^H - F_{cal}^H)}{(\sigma F_{obs}^H)^2} \exp(-2\pi i H \cdot r) + 1 = 0$$

This leads to
$$\ln(\rho_i) = Z + 1 - \frac{2Z\lambda}{M} \times \sum \frac{(F_{obs}^H - F_{cal}^H)}{(\sigma F_{obs}^H)^2} \exp(-2\pi i H \cdot r)$$

and then
$$\rho_i = \exp\left[ Z + 1 - \frac{2Z\lambda}{M} \times \sum \frac{(F_{obs}^H - F_{cal}^H)}{(\sigma F_{obs}^H)^2} \exp(-2\pi i H \cdot r) \right]$$

In the iterative process of MEM, the value of $\rho_i$ in the $(n+1)$th iterative cycle, $\rho_i^{n+1}$, is derived from the electron density image of the $n$th cycle using the above equation, namely

$$\rho_i^{n+1} = \exp\left[ Z + 1 - \frac{2Z\lambda}{M} \times \sum \frac{(F_{obs}^H - F_{cal,n}^H)}{(\sigma F_{obs}^H)^2} \exp(-2\pi i H \cdot r) \right]$$

where $F_{cal,n}^H$ is the calculated structure factor obtained by Fourier transforming the electron density image of the $n$th iterative cycle.

Before the iterative process of MEM, a positive initial value is assigned to $\rho_i$. In most cases, a uniform positive value is assigned to each grid as the initial value, which is known as the uniform model. As $\rho_i$ is an exponential function, it is always positive in the iterative process.

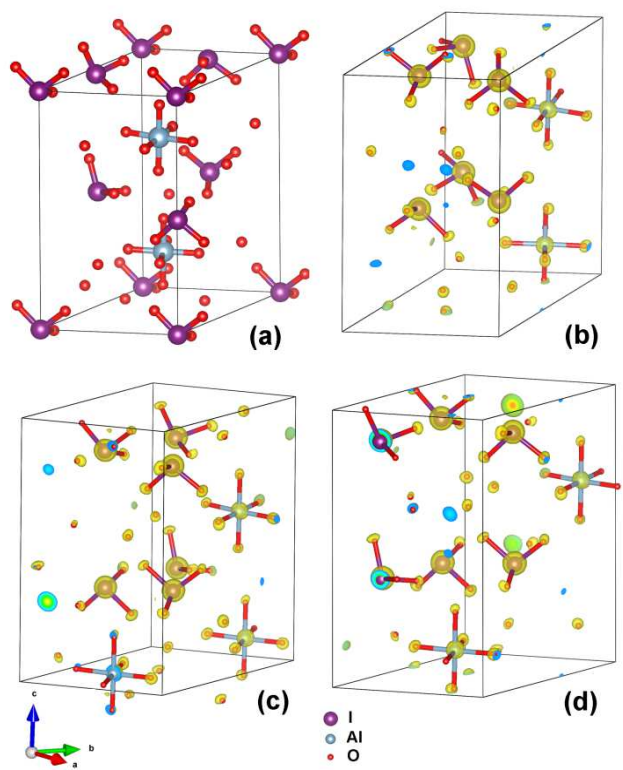

**Figure S1.** (a) The atomic structure model of $Al(IO_3)_3(H_2O)_8$ and the electron density images reconstructed with (b) the algorithm based on Tian1, (c) the error reduction and (d) charge flipping algorithms. The original atomic structure model was superimposed upon the electron density images to show the consistence between them. The atomic coordinates in the structure model are translational shifted and/or inverted to make a direct comparison with the electron density images in (b), (c) and (d).